\documentclass[%
reprint,
10pt,
superscriptaddress,
nofootinbib,
 amsmath,amssymb,
 aps,
 prl,
 twocolumn,
]{revtex4-2}

\usepackage{graphicx}
\usepackage{dcolumn}
\usepackage{bm}

\usepackage{physics}     
\usepackage{xcolor}      
\usepackage{color,soul}
\usepackage{hyperref}    
\usepackage{float}       
\usepackage{transparent} 
\usepackage{layouts}     

\begin{document}
\newcommand{\mycite}[1]{\citejournal{#1},{\bfseries \citevolume{#1}}, \citepages{#1}, \citeyear{#1}}

\newcommand{\adj}[1]{\hat{#1}^{\dagger}}
\newcommand{\oper}[1]{\hat{#1}^{}}
\newcommand{\ssOper}[2]{\hat{#1}^{#2}}
\newcommand{\rhoSS}{\oper{\rho}_{ss}}
\newcommand{\rhoSSLL}{\ssOper{\rho}{ss}_{ll}}
\newcommand{\Liou}{\oper{\mathcal{L}}}
\newcommand{\steadyInten}[2]{I_{#1}^{#2}}
\newcommand{\rhoZ}{\oper{\rho}(z)}
\newcommand{\rhoNull}{\oper{\rho}_{in}}
\newcommand{\rhoF}{\oper{\rho}_{f}}
\newcommand{\zm}{\ssOper{\rho}{0}}
\newcommand{\ULiou}{\oper{U}_{\hat{\mathcal{L}}}}

\newcommand{\NPhotSubSpace}[1]{\ssOper{\rho}{(#1)}}
\newcommand{\NPhotInten}[2]{I^{(#1)}(#2) }
\newcommand{\CL}{C_{l_1}}
\newcommand{\Heff}{\hat{H}_{\text{eff}}}
\newcommand{\iket}[1]{\ket*{#1}}
\newcommand{\ibra}[1]{\bra*{#1}}
\newcommand{\iketbra}[2]{\iket{#1}\! \! \! \ibra{#2}}
\newcommand{\ikket}[1]{ |  #1 \rangle \! \rangle}

\newcommand{\cn}[1]{\colorbox{orange}{citation needed: #1}}
\newcommand{\rn}[1]{\colorbox{red}{revision needed: #1}}
\newcommand{\todo}[1]{\colorbox{cyan}{todo: #1}}
\newcommand{\done}[1]{\colorbox{green}{done: #1}}

\newcommand*{\shortautoref}[1]{%
  \begingroup
    \def\sectionautorefname{Sec.}%
    \def\subsectionautorefname{Subsec.}%
    \def\figureautorefname{Fig.}%
    \def\equationautorefname{Eq.}%
    \def\chapterautorefname{Ch.}%
    \def\tableautorefname{Tab.}%
    \autoref{#1}%
  \endgroup
}

\title[Non-Hermitian topological filters]{
Non-Hermitian topological filters
}

\author{Vinzenz Zimmermann}
\email{vinzenz.zimmermann@ucf.edu}
\affiliation{CREOL, The College of Optics and Photonics, University of Central Florida, Orlando, FL 32816, USA}

\author{Amin Hashemi}
\affiliation{CREOL, The College of Optics and Photonics, University of Central Florida, Orlando, FL 32816, USA}

\author{Kurt Busch}
\affiliation{Max-Born-Institut, Max-Born-Strasse 2A, 12489 Berlin, Germany}
\affiliation{Humboldt-Universität zu Berlin, Institut für Physik, AG Theoretische Optik \& Photonik, Berlin, Germany}

\author{Andrea Blanco-Redondo}
\affiliation{CREOL, The College of Optics and Photonics, University of Central Florida, Orlando, FL 32816, USA}

\author{Armando Perez-Leija}
\email{armando.leija@slu.edu}
\affiliation{Department of Electrical and Computer Engineering, Saint Louis University, St. Louis, Missouri 63103, USA}

\date{\today}

\begin{abstract}
We introduce a non-Hermitian photonic filter that harnesses dissipation to selectively isolate a desired topological state. In science and engineering, dissipation is often used to filter incoherent waves, producing a pure coherent output. Here, we apply this principle to topological states, creating a linear filter that effectively isolates a specific topological state regardless of the initial input's coherence properties.
This approach creates a dissipation-free topological subspace, where the desired states are preserved and their topological protection is enhanced. Our work provides a versatile and simple method for topological state selection, opening the door to new applications in integrated topological photonics.
\end{abstract}

\keywords{coherence, non-Hermicity, topological protection}
\maketitle

\section{I. Introduction}
The distillation of coherent waves from incoherent sources via dissipative processes has a well-established history in physics.
 In this context, perhaps the oldest and most famous example is the double-slit interference experiment performed by Thomas Young in 1801 \cite{young_i_1804}. In this experiment, Young was able to filter a bundle of spatially coherent light modes from a highly incoherent source by discriminating most of the light using pinholes in an otherwise opaque plate. Thereafter, the notion of dissipation-assisted coherence distillation has recurrently appeared in classical and quantum physics in many different contexts \cite{mi_stable_2024, selim_selective_2025}. For instance, in the framework of open quantum systems, an appropriate manipulation of loss mechanisms, that is, a careful design of the interaction between the system-of-interest and the environment, can drive the former into predefined pure states \cite{diehl_quantum_2008, hashemi_observation_2025}. Similarly, engineered dissipation can be used to produce an interesting variety of strongly correlated quantum states that are useful for universal quantum computing \cite{verstraete_quantum_2009}. Quite remarkably, dissipation can also be harnessed in systems with rather fragile quantum dynamics such as trapped ions, to deterministically produce and stabilize maximally entangled Bell states \cite{lin_dissipative_2013}.\\
 \indent
 In essence, these works demonstrated that by including dissipation at precise sites of certain dynamical systems, one can establish the conditions for the existence of dissipation-free coherent states. 
A defining characteristic of these states is that their eigenvalues vanish. In general, these states are termed dark states, or simply zero modes, and they have attracted significant attention because they represent a naturally protected, coherent, and stable subspace within an otherwise dissipative system \cite{ge_symmetry-protected_2017}. Physically, this means that, as excitations (formed by arbitrary combinations of allowed states) propagate, the system acts as a filter where most constituent states decay due to dissipation, while the dissipation-free ones remain intact.\\
\indent
Due to this potential, zero modes have been investigated in a variety of contexts ranging from topological insulators \cite{hasan_colloquium_2010, qi_topological_2011}, 
Majorana fermions \cite{alicea_new_2012, reiserer_cavity-based_2015, elliott_colloquium_2015}, and high-dimensional multipole insulators \cite{benalcazar_quantized_2017, peterson_quantized_2018, el_hassan_corner_2019}. A detailed analysis on the coupling requirements between zero modes and bulk states to produce spectrally embedded zero modes in the bulk continuum, mid-gap zero modes, and to restore the “zeroness” of disorder-shifted topological states can be found in Ref. \cite{rivero_robust_2023}.\\
\indent
In optics, zero modes have enabled important advances in non-Hermitian single-mode lasers \cite{ge_symmetry-protected_2017, zhao_topological_2018, poli_selective_2015, peng_paritytime-symmetric_2014}. 
In photonic crystals, zero modes manifest as highly localized defect modes \cite{mertens_tunable_2005}, and they allow for the creation of ultra-compact optical filters and high-quality factor cavities \cite{kuzmiak_localized_1998, painter_defect_1999}. In fact, zero modes can be strategically introduced and tuned in photonic crystals by controlling the geometry and symmetry of the defects \cite{villeneuve_microcavities_1996}.


In parallel to progress in research and applications of zero modes and photonic dark states, there has been a great deal of interest in photonic topological insulators (PTIs), which are optical systems that exhibit exceptionally robust light transport \cite{hafezi_robust_2011, khanikaev_photonic_2013,  tschernig_topological_2021, ren_topologically_2022, tschernig_topological_2022,  vega_topological_2023, pan_harnessing_2024}. Importantly, transport robustness in PTIs is associated with the existence of topological states whose propagation eigenvalues fall within a relatively wide spectral gap, isolated from the eigenvalues of non-topological states \cite{kitagawa_topological_2010,  rechtsman_photonic_2013, plotnik_observation_2014}. This implies that in the presence of certain stationary perturbations, topological states undergo negligible coupling with non-topological states, thereby preventing dissipation into the bulk of the system \cite{bandres_topological_2016, stutzer_photonic_2018}. Notwithstanding these notable properties, topological states are in general susceptible to environmental dissipation. It is therefore of practical relevance to identify dissipation conditions under which PTIs still support their topological states while the non-topological ones become dissipative. 

In this work, we combine the idea of dissipation assisted filtering with the properties of photonic zero modes to introduce a non-Hermitian photonic filter for topological states. Our approach is based on the unique property of certain photonic topological insulators (PTIs) that feature a single topological state whose propagation constant is zero. Hence, by adding engineered dissipation at precise sites within the PTI, the supported zero mode  becomes the system's unique steady state. This creates a highly versatile linear filter that effectively isolates the desired topological state, regardless of the coherence properties of the input field.

\indent
Indeed, similar ideas have been investigated in the fully coherent regime using non-Hermitian arrays of microwave resonators \cite{poli_selective_2015}. 
Here, we expand on these ideas to create a photonic topological filter capable of operating on any type of excitation, whether coherent, partially coherent, or fully incoherent. To demonstrate this capability, we evaluate the filter performance across a range of input fields and analyze the proposed structures in the Liouville (or correlation) space.

The paper is structured as follows. Section II provides a detailed theoretical background for zero modes in photonic waveguide systems. Building on these concepts, Section III introduces a specific PTI that when combined with engineered dissipation, results in a non-Hermitian filter whose steady state is precisely the topological mode. Finally, we summarize our work and conclude in Section IV.

\section{II. Zero Modes in Hermitian and non-Hermitian waveguide systems}
To illustrate the basic idea of non-Hermitian photonic filters, we first discuss the emergence of photonic zero modes in both lossless and dissipative one-dimensional systems of evanescently coupled waveguides.
To do so, we consider a system comprising three identical lossless coupled waveguides, each waveguide exhibiting a step refractive index profile that features a single fundamental mode. Under these premises, the tight-binding Hamiltonian governing this system is 
\begin{align}\label{eq:Hamiltonian}
 \hat{H}=\begin{pmatrix}
0 & \kappa & 0\\
\kappa & 0 & \kappa\\
0 & \kappa & 0
\end{pmatrix},
\end{align}
where $\kappa$ represents the coupling strength between adjacent waveguides. Direct diagonalization of $\hat{H}$ yields three eigenmodes, including $\ket{\psi_0} = (1,0,-1)^{T}/\sqrt{2}$ whose eigenvalue $\lambda_{0}=0$.\\
\begin{figure}[t!]
    \centering
    \includegraphics[width =  \columnwidth]{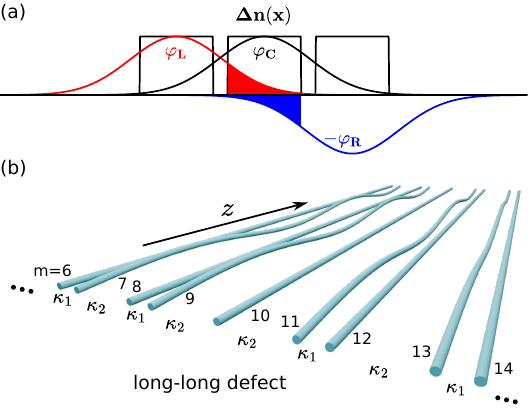}
    \caption{a) Destructive interference of the evanescent amplitudes of $\varphi_L$, and $-\varphi_R$ occurring in the central waveguide.
    Black curves represent the refractive index profile defining the waveguide trimer. $\Delta n(x)$ is the refractive-index profile of the central waveguide. The blue and red shaded areas represent the terms $\varphi_{L}^{*}(x)\Delta n$ and $\varphi_{R}^{*}(x)\Delta n(x)$, respectively. Hence, is clear that the integral defining the simultaneous coupling $\kappa_S$ in Eq. \ref{eq:CoupledEqn} is identically zero. 
    b) Central section of a long-long defective SSH PTI lattice with engineered dissipation in the odd parity sublattice. The defect waveguide is located right at the center of the array $(m=10)$. Losses are here realized by sinusoidal modulations of the waveguides along the propagation direction $z$.}
    \label{fig:SSH_LL_Defect_sketch}
\end{figure}
To elucidate the physical characteristics of this zero mode, let us denote the individual modes supported by the left, central, and right waveguides as $\varphi_L$, $\varphi_C$, and $\varphi_R$, respectively, see Fig~(\ref{fig:SSH_LL_Defect_sketch}-a). We then compute the simultaneous coupling $\kappa_S$ between $\varphi_C$ and both $\varphi_L$ and $\varphi_R$. According to coupled-mode theory \cite{yariv_coupled-mode_1973}, $\kappa_S$ is proportional to the integral 
\begin{align}\label{eq:CoupledEqn}
\kappa_S \propto \int (\varphi_{L}^{*}(x) +\varphi_{R}^{*}(x))\Delta n(x) \varphi_{C}(x) dx,
\end{align}
where $\Delta n(x)$ represents the refractive index profile of the central waveguide. As $\Delta n(x)$ is considered to be a symmetric function, and assuming that $\varphi_{L}^{*}(x)$ and $\varphi_{R}^{*}(x)$ have the same amplitude $1/\sqrt{2}$ and a $\pi$-phase difference, as indicated by the amplitudes of $\ket{\psi_0}$, it is clear that the sum $(\varphi_{L}^{*}(x) +\varphi_{R}^{*}(x))\Delta n =0$. Consequently, the simultaneous coupling $\kappa_S =0$. Physically, this means that the evanescent amplitudes of $\varphi_{L}(x)$ and $-\varphi_{R}(x)$ destructively interfere at the central waveguide, preventing their simultaneous coupling to the latter. These effects are visually illustrated in Fig~(\ref{fig:SSH_LL_Defect_sketch}-a). In similar fashion, we can show that waveguide trimers involving dissimilar couplings $(\kappa_1, \, \kappa_2)$ also feature a zero-eigenvalue mode. The difference is that the amplitudes in the outer waveguides of such a zero mode are no longer equal, but the $\pi$-phase difference persists, ensuring the cancellation of the evanescent fields within the central waveguide. \\ 
\indent
Given that the zero mode does not have any physical support to excite the central waveguide, it is rather evident that the introduction of dissipation in the central waveguide will not affect its behavior. This can be confirmed by solving the corresponding eigenvalue problem for the effective, non-Hermitian Hamiltonian 
\begin{align}\label{eq:Hamiltonian_NHH}
 \hat{H}_{\text{eff}}=\begin{pmatrix}
0 & \kappa & 0\\
\kappa & i\gamma & \kappa\\
0 & \kappa & 0
\end{pmatrix},
\end{align}
where $\gamma$ represents the dissipation rate in the central waveguide. By doing so, we find that $\ket{\psi_0} = (1,0,-1)^{T}/\sqrt{2}$ is also an eigenmode of $\Heff$ with eigenvalue zero. Therefore, the zero mode remains invariant under the addition of losses to the central waveguide. The cancellation of evanescent field amplitudes in the intermediate waveguide is the mechanism that prevents light from coupling into this site.\\

\section{III. Non-Hermitian topological filters}
As shown in the previous section, photonic lattice systems comprising an odd number of waveguides can be engineered to support zero modes. These zero modes are antisymmetric stationary solutions characterized by evanescent amplitude cancellation at even-numbered waveguides, which prevents their coupling to intermediate, possibly dissipative, waveguides.
In fact, a variety of Hermitian PTIs host topological states exhibiting zero eigenvalues \cite{ryu_topological_2002, guzman-silva_experimental_2014, califrer_proximity-induced_2023}, a class typified by the Su-Schrieffer-Heeger (SSH) topological lattice \cite{su_solitons_1979}. 
Hence, for simplicity, and without loss of generality, we perform our analysis using the SSH PTI.
Although this paper focuses primarily on PTIs, we emphasize that the proposed approach applies to other lattice configurations, provided they support zero modes.\\ 
\indent
In optics, topological SSH systems can be implemented using one-dimensional waveguide systems designed to have alternating short and long gaps between adjacent waveguides that tailor their coupling strength. In addition, a defect waveguide that breaks the short-long coupling sequence is introduced in the middle of the array to implement the topological interface \cite{blanco-redondo_topological_2016, wang_topologically_2019}, Fig.~(\ref{fig:SSH_LL_Defect_sketch}-b).\\
\indent
For our analysis, we consider a non-Hermitian SSH system formed by $M=21$ waveguides with alternating coupling coefficients $\kappa_{1}=2\text{ cm}^{-1}$, and $\kappa_{2}=1\text{ cm}^{-1}$. Dissipation is included in odd-numbered waveguides with a dissipation rate $\gamma=1\text{ cm}^{-1}$. These coupling and dissipation parameters are feasible using current integrated waveguide technology, e.g. femtosecond-laser-writing techniques \cite{szameit_control_2007}, or silicon nanowires \cite{blanco-redondo_topological_2018}. Using the former approach, a non-Hermitian SSH system can be implemented by writing non-equidistant straight waveguides, while dissipation can be obtained by sinusoidally modulating every second waveguide as illustrated in Fig.~(\ref{fig:SSH_LL_Defect_sketch}-b) \cite{eichelkraut_mobility_2013}.

\begin{figure}
    \centering
    \includegraphics[width =  \columnwidth]{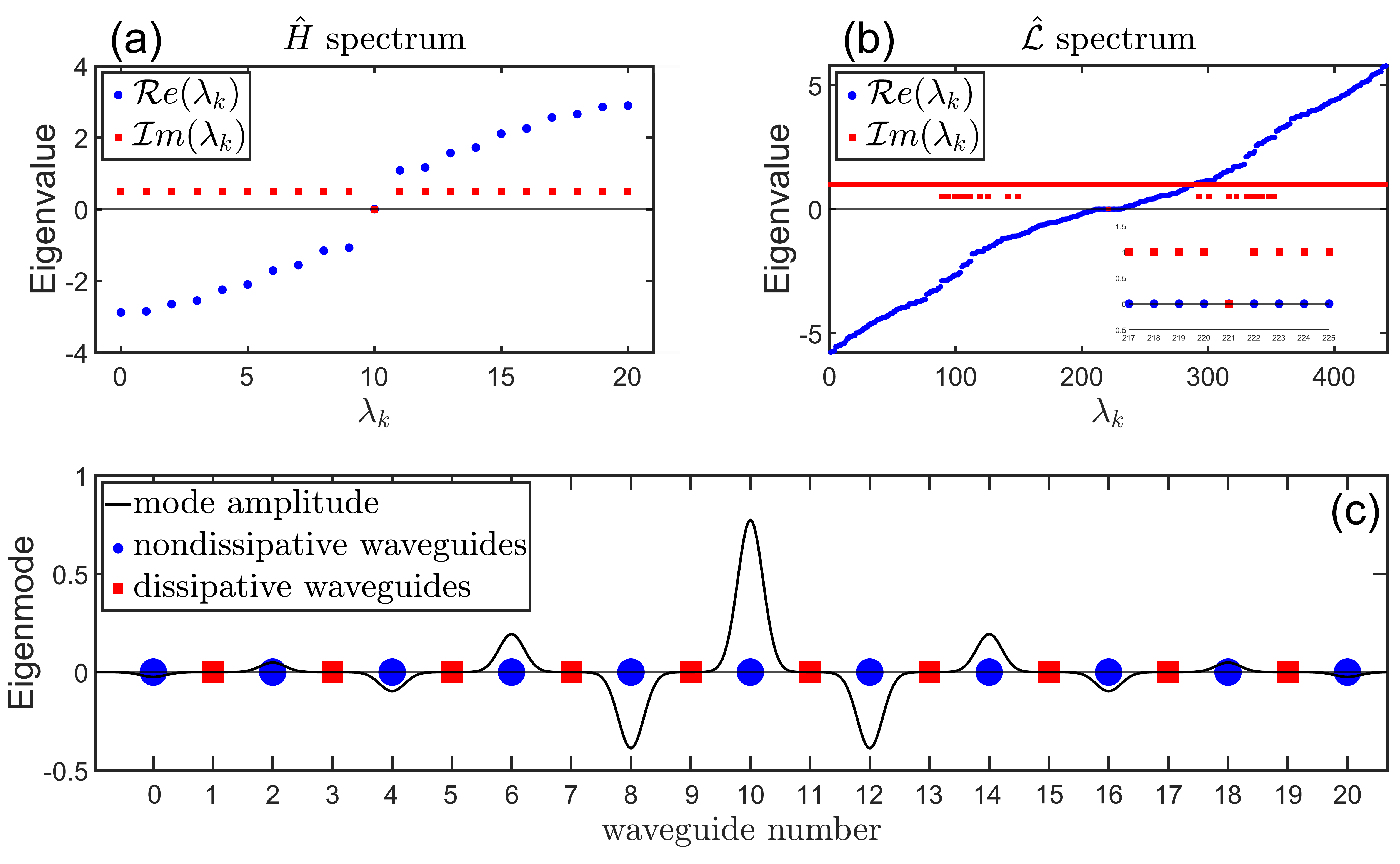}
    \caption{a) Eigenvalue spectrum of a non-Hermitian SSH system comprising $M=21$ waveguides with coupling coefficients $\kappa_1 = 2\text{ cm}^{-1}$, $\kappa_2 = 1 \text{ cm}^{-1}$, and dissipation rate $\gamma=1\text{ cm}^{-1}$. b) Liouville (correlation) spectrum. Inset in b) is a close-up of the central part of the spectrum showing that the eigenvalue for the topological mode 221st eigenvalue is purely real. c) Topological zero-eigenvalue mode and its amplitudes populating the non-Hermitian SSH lattice. }
    \label{fig:SpectrumMode}
\end{figure}

The evolution of coherent light in the proposed non-Hermitian SSH waveguide array is governed by the set of differential equations
\begin{align}\label{eq:Sch}
-i \frac{d}{dz}E_m(z) =  \sum_{n=0}^{M}\left[\Heff\right]_{m,n}E_n(z),
\end{align}
where $[\Heff ]_{m,n}$ are the matrix elements of the non-Hermitian Hamiltonian with diagonal terms $[\Heff ]_{m,m}=i\gamma(1+(-1)^{m+1})/2$, and upper diagonal elements $[\Heff ]_{m,m+1} = \left[\frac{\kappa_1}{2}(1+(-1)^{m}) +\frac{\kappa_2}{2}(1+(-1)^{m+1})\right]$, for $m=0,...,(\frac{M-1}{2})-1$, and $[\Heff ]_{m,m+1} = \left[\frac{\kappa_2}{2}(1+(-1)^{m}) +\frac{\kappa_1}{2}(1+(-1)^{m+1})\right]$, for $m=\frac{M-1}{2},...,M-2$. Additionally, the lower diagonal elements are given by the H.c. elements of $[\Heff ]_{m,n}$.
The spectrum for this system with $M=21$ waveguides is depicted in Fig.~(\ref{fig:SpectrumMode}-a). The eigenvalue $\lambda_{0} = 0$ corresponds to the topological state $\ket{\psi_{0}}$, whose amplitudes vanish at dissipative sites, while at consecutive lossless waveguides it features a $\pi$ phase difference, see  Fig.~(\ref{fig:SpectrumMode}-c).
Clearly, this SSH system preserves the topological zero mode, whereas all remaining eigenmodes become dissipative, as indicated by their complex eigenvalues.

In general, spatial partially coherent light can be represented as a large ensemble of light fields with random amplitudes and phases. Due to this inherent randomness, Eq.~\eqref{eq:Sch} is insufficient for describing the light dynamics, and the analysis must be performed within the field correlation space, also known as Liouville space.\\
\indent
The behavior of a partially coherent light under the scalar wavefront approximation is described by the mutual intensity function $\rho(z)$ comprising the $M^2$ first-order correlation functions $\rho(z)=\left(\rho_{0,0}(z),\hdots,\rho_{1,M-1}(z),\rho_{2,1}(z),\hdots,\rho_{M-1, M-1}(z)\right)^{T}$, where $\rho_{m,n}(z)=\langle E_{m}(z)E_{n}^{*}(z)\rangle$ involves the average of the light amplitudes in waveguides $m$ and $n$, and superscript $T$ denotes transpose.
For linear waveguide tight-binding lattices, the evolution of $\rho(z)$ is governed by Liouville’s equation
\begin{align}
-i\frac{d\rho(z)}{dz} =  \hat{\mathcal{L}}\rho(z) \, .
\label{eq:LiouvilleEq}
\end{align}
Here, $\hat{\mathcal{L}}=\Heff \otimes \hat{I}-\hat{I} \otimes \Heff^{\dagger}$ is the Liouvillian, $\Heff$ is the non-Hermitian Hamiltonian described in Eq.~\eqref{eq:Sch}, $\Heff^{\dagger}$ is the adjoint of $\Heff$, and $\hat{I}$ is the identity matrix of the same dimensions.\\
\indent

In the same way as diagonalizing the effective Hamiltonian $\Heff$ reveals the properties of dissipative systems within the fully coherent subspace, exploring the Liouvillian spectrum yields essential information, specifically decay rates and steady states, about the system when excited by partially coherent light.\\
\indent
To compute the Liouvillian spectrum, we exploit the fact that the non-Hermitian Hamiltonian for the proposed systems is symmetric, that is, $\Heff^{\dagger}=\Heff^{*}$. This implies that the right eigenvectors $\ket{\phi_k}$ of $\Heff$ (with eigenvalues $\lambda_k$) obey an analogous relationship for their complex conjugates. That is, the complex conjugate states $\ket{\phi_{k}^{*}}$  satisfy the equation $\Heff^{*}\ket{\phi_{k}^{*}}=\left( \Heff\ket{\phi_k}\right)^{*}=\lambda_{k}^{*}\ket{\phi_{k}^{*}}$.
Hence, assuming that  the Kronecker product $\ket{\phi_m}\ket{\phi_{n}^{*}}$ is an eigenvector of $\hat{\mathcal{L}}$, we find the Liouvillian eigenvalues
\begin{align}\nonumber
\hat{\mathcal{L}}\ket{\phi_m}\ket{\phi_{n}^{*}}&=\Heff\ket{\phi_m}\otimes\hat{I}\ket{\phi_{n}^{*}}-\hat{I}\ket{\phi_m}\otimes\Heff^{*}\ket{\phi_{n}^{*}}\\
&=\left(\lambda_{m}-\lambda_{n}^{*}\right)\ket{\phi_m}\ket{\phi_{n}^{*}}.
\end{align}
That is, the $\hat{\mathcal{L}}$ eigenvalues are given by the difference of the $\Heff$ eigenvalues and their complex conjugate. Importantly, this fundamental subtractive property of the Liouville eigenvalues removes the topological gap, see Fig.~(\ref{fig:SpectrumMode}-b). However, as the eigenvalue spectrum of the non-Hermitian SSH contains $(M-1)$ complex eigenvalues with the same imaginary part, e.g., Fig.~(\ref{fig:SpectrumMode}-a), all combinations $\left(\lambda_{m}-\lambda_{n}^{*}\right)$ involving such complex eigenvalues are also complex, see Fig. ~(\ref{fig:SpectrumMode}-b). This means that in correlation space the non-Hermitian SSH supports a single zero mode (see inset in Fig. ~(\ref{fig:SpectrumMode}-b)), which is given by the Kronecker product of the $\Heff$ topological mode and its complex conjugate, $\rho_{0}=\ket{\psi_{0}}\ket{\psi_{0}^{*}}$. 
\\
\indent
As noted above, spatial partially coherent light generally exhibits random amplitudes and phases. Due to this inherent randomness, the system is assured to be excited by light configurations that span all possible mode distributions, thereby guaranteeing the excitation of the topological zero mode.
Crucially, since the topological zero mode is the only one immune to dissipation, all other excited modes will vanish over time, causing the system to relax into a pure topological steady state.
\begin{figure}[t!]
    \centering
    \includegraphics[width = \columnwidth]{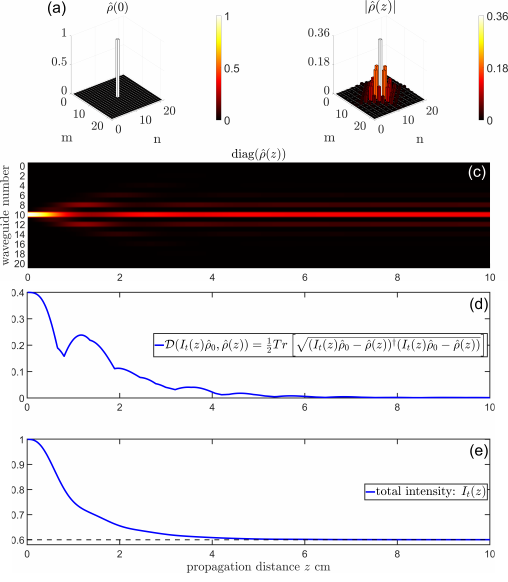}
    \caption{Coherent light dynamics in a non-Hermitian topological SSH filter. (a) Coherent single site excitation $\hat{\rho}(0)$ at the $m= 10$th waveguide. (b) Coherent topological steady state $\hat{\rho}_{0}(z)$ at the output of the array. (c) Diagonal elements of the correlation matrix $\rhoZ$ representing the evolution of the waveguide intensities $I_m(z)$. (d) Trace distance as a function of the propagation coordinate $z$. (e) Evolution of the total intensity $I_{t}(z)$.}
\label{fig:CoherentEvolution}
\end{figure} 
\begin{figure*}[t!]
    \centering
    \includegraphics[width = \textwidth]{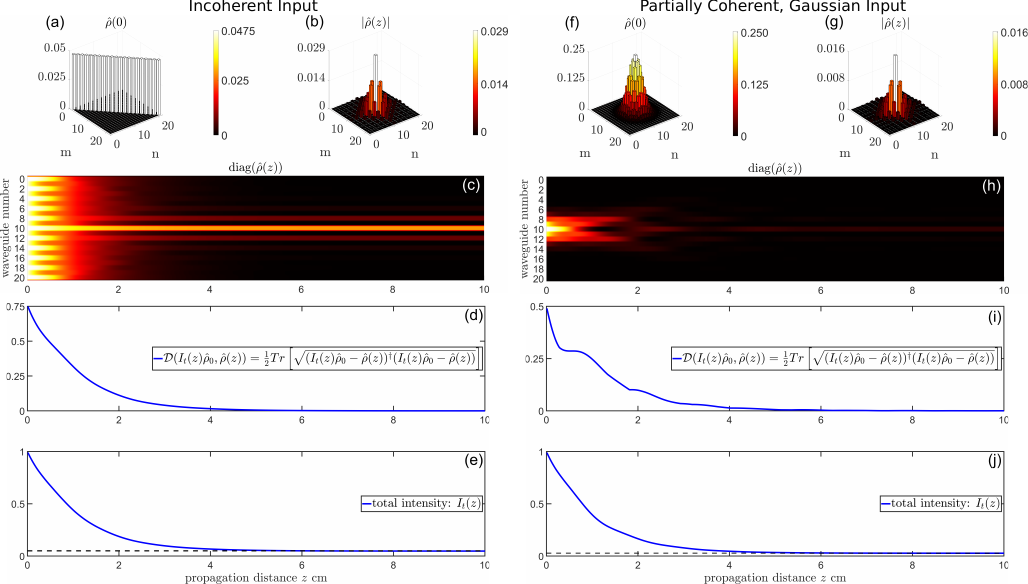}
    \caption{Incoherent (left) and partially coherent (right) light dynamics in a non-Hermitian topological SSH filter. (a), (f) Input correlation matrix $\hat{\rho}(0)$ representing a fully incoherent excitation of the entire waveguide system and a partially coherent excitation with Gaussian profile centered at the $m = 10$th waveguide. (b), (g) Filtered coherent topological steady state $\hat{\rho}_{0}(z)$ at the output of the system. (c), (h) Diagonal elements of the correlation matrix $\rhoZ$ representing the evolution of the waveguide intensities $I_m(z)$. The colormaps were normalized independently. (d), (i) Trace distance as a function of the propagation coordinate $z$. (e), (j) Evolution of the total intensity $I_{t}(z)$.}
    \label{fig:IncoherentEvolution}
\end{figure*} 

To illustrate these concepts, we consider the evolution of light with different degrees of coherence. As a coherent input, we consider the excitation of the central site, which in correlation space corresponds to $\oper{\rho}(0)=\delta_{10,10}$ as shown in Figs.~(\ref{fig:CoherentEvolution}-a,c). Notice, in what follows the ``hat" symbol $\hat{\cdot}$ indicates that the vectors have been expressed in matrix form. 
Under this single-waveguide excitation, light tends to couple to neighboring waveguides but dissipation prevents it from spreading by fading the amplitudes of non-topological modes. This is clearly seen in the intensity dynamics depicted in Fig.~(\ref{fig:CoherentEvolution}-c). To determine the propagation distance at which the system filters the topological mode, we compute the trace distance between the propagating $\hat{\rho}(z)$ and the topological $\hat{\rho}_{0}$  correlation matrices,
\begin{align}
\mathcal{D}(\hat{\rho}_{0},\hat{\rho}(z))=\frac{1}{2}Tr\left[\sqrt{(\hat{\rho}_{0}-\hat{\rho}(z))^{\dagger}(\hat{\rho}_{0}-\hat{\rho}(z))}\right].
\label{eq:TD}
\end{align}

In quantum information, the trace distance is a metric used to measure the distinguishability between two quantum states \cite{nielsen_quantum_2010}. In the present context, we use it to quantify  the similarity of the two light states represented by correlation matrices $\hat{\rho}(z)$ and $\hat{\rho}_{0}$. 
In general, the trace distance is defined for normalized correlation matrices and is bounded in the range
$0\le\mathcal{D}(\hat{\rho}_{0},\hat{\rho}(z))\le1$, with $\mathcal{D}(\hat{\rho}_{0},\hat{\rho}(z))=0$ if and only if the matrices are identical, $\hat{\rho}_{0}=\hat{\rho}(z)$.
Hence, to take into account the fact that the total intensity $I_{t}(z)$ decreases along $z$, in Eq.~\eqref{eq:TD} we substitute $\hat{\rho}_{0}$ by $I_{t}(z)\hat{\rho}_{0}$.

Fig.~(\ref{fig:CoherentEvolution}-d) shows the trace distance as a function of $z$. In the interval $z\in[4,6]$ cm, the trace distance becomes negligible, $\mathcal{D}(I_{t}(z)\hat{\rho}_{0},\hat{\rho}(z))\approx0$, indicating that the system has filtered out all non-topological modes. At $z=6$ cm, the system reaches the topological steady state, $\hat{\rho}(z)=\hat{\rho}_{0}$. This is further supported by inspecting the evolution of the total intensity $I_{t}(z)$ shown in Fig.~(\ref{fig:CoherentEvolution}-e). This curve reveals that the system dissipates $40\%$ of the input energy before reaching the steady state, with the total intensity remaining constant from $z=5$ cm onward.\\ 
\indent
We now turn our attention to filtering the topological mode from fully incoherent  and partially coherent light. For the incoherent input, the correlation matrix is diagonal and characterized by a uniformly weighted superposition of mutually uncorrelated modes coupled to the entire waveguide system, $\hat{\rho}(0)=\hat{I}/M$, as illustrated in Fig.~(\ref{fig:IncoherentEvolution}-a). 
When such uncorrelated light enters the system, the correlation terms involving only dissipative waveguides are the first to decay as they experience a doubled dissipation rate $(2\gamma)$, the explanation for this effect is given below.
In Fig.~(\ref{fig:IncoherentEvolution}-d) we show the computed trace distance. 
In contrast to the fully coherent input scenario, where the trace distance exhibits oscillations while decreasing, in the present case it decreases monotonically, becoming zero at $z\approx4$ cm.
These results clearly show that after a propagation distance of about $4 \text{ cm}$, the correlation matrices for coherent and incoherent excitations become virtually indistinguishable and the system reaches the topological steady state after $z=6$ cm. That is, regardless of the amount of coherence in the initial state, the system always evolves towards the topological steady state.
However, as about half ($(M-1)/2$) of the waveguides in our non-Hermitian filter are dissipative and the fully incoherent light input spans the entire array, only $4\%$ of the energy remains in the topological steady state. In other words, since roughly half of the energy is coupled into dissipative waveguides, the filtering efficiency is much lower than in the coherent excitation case.
The situation improves when considering a fully incoherent excitation of the three non-dissipative central waveguides $(8\text{th}, 10\text{th}, 12\text{th})$ with the same intensity. Our numerical experiments showed an outcome similar to the previous incoherent case, that is, the topological steady state becomes isolated within the interval $z\in[4,6]$ cm and maintains $30\%$ of the total intensity. Due to the similarity with the previous case, these results are not shown.\\ 
\indent
To benchmark the performance of our system for partially coherent light, we consider an initial field described by the Gaussian correlation profile $\hat{\rho}(0) = \left(\sqrt{2\pi} \sigma \right) ^{-1} \exp{-((m-m_{10})^2+(n-n_{10})^2)/2\sigma}$ with spatial width $\sigma = 5$ as shown in Figs.~(\ref{fig:IncoherentEvolution}-f,h). 
Although the initial trace distance $\mathcal{D}(\hat{\rho}_{0},\hat{\rho}(0))$ indicates a high similarity to the targeted topological zero mode compared to the incoherent input, see Fig.~(\ref{fig:IncoherentEvolution}-i), the filtering performance is quantitatively poor. This is demonstrated by the low energy yield at the output where only 2.3\% of the initial energy remains in the topological steady state. The system's performance is critically dependent on the spatial profile width $\sigma$: decreasing $\sigma$ significantly improves the filtering, leading to results comparable to the coherent scenario (Fig.~(\ref{fig:CoherentEvolution})), whereas increasing $\sigma$ causes further degradation.\\
\indent
These results conclusively demonstrate that the topological steady state is reached irrespective of the input light's coherence properties or spatial profile. This finding underscores the significant potential of non-Hermitian filtering as a robust mechanism to realize these topologically protected photonic dark states, eliminating the need to precisely engineer the initial input state $\hat{\rho}_{0}$.\\
\begin{figure}[t!]
    \centering
    \includegraphics[width = \columnwidth]{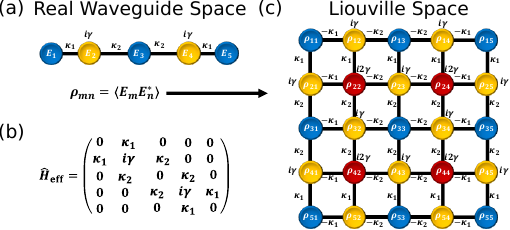}
    \caption{(a) Non-Hermitian waveguide system in real space. $\kappa_{1,2}$ are the coupling coefficients and $\gamma$ is the dissipation rate. (b) Non-Hermitian Hamiltonian that governs light dynamics in the waveguide array in (a).
    (c) Liouville (correlation) space corresponding to the waveguide system (a). The Liouvillian for the system pictorially shown in (c) is obtained using the expression $\hat{\mathcal{L}}=\Heff\otimes\hat{I}-\hat{I}\otimes\Heff^{\dagger}$, where $\Heff$ is the non-Hermitian Hamiltonian depicted in (b).}
    \label{fig:RealLiouville}
\end{figure} 
Before concluding, we briefly discuss why some coherence terms undergo a doubled dissipation rate, as pointed out above. To do so, it is elucidating to describe the correspondence between the real waveguide space and the Liouville (correlation) space. For this purpose, in Fig.~(\ref{fig:RealLiouville}) we show the actual and correlation space for a non-Hermitian waveguide system comprising $M=5$ sites with coupling coefficients $\kappa_{1,2}$, and dissipation rate $\gamma$. Notably, in Liouville space, the correlation terms exhibit negative and positive couplings in the horizontal and vertical directions, respectively. Furthermore, correlations involving fields in lossless waveguides do not present any dissipation (e.g., $\rho_{35}$). However, correlations involving the field in a lossless waveguide and the field in a dissipative waveguide (e.g., $\rho_{25}$) undergo the same amount of dissipation as in real space $(\gamma)$, while correlations between fields reaching either two dissipative sites (e.g., $\rho_{25}$) or the same dissipative site (e.g., $\rho_{44}$) exhibit twice the dissipation rate $(2\gamma)$. This implies that partially coherent light characterized by correlations involving at least one dissipative waveguide will unavoidably decay. 
These observations apply broadly to non-Hermitian Su-Schrieffer-Heeger (SSH) systems with an arbitrary number of waveguides. Furthermore, the correlation picture clearly explains the pronounced drop in filtering performance for partially coherent light (e.g. Fig.~(\ref{fig:IncoherentEvolution}-f-i) compared to the fully incoherent scenario (Fig.~(\ref{fig:IncoherentEvolution}-a-e)).
Simply put, partially coherent light is more prone to degradation regardless of the higher similarity between the input and the target state. Generalizing these results will provide valuable guidelines for designing efficient initial excitations in non-Hermitian systems.\\

\section{IV. Conclusions}
We have demonstrated that non-Hermitian photonic lattices can function as topological filters. This is achieved by strategically introducing losses into a specific waveguide subspace, which causes all unwanted, unprotected modes to dissipate while preserving the desired topologically protected mode. This unique filtering behavior ensures that the system always produces the same topologically protected output, regardless of the coherence and spatial profile of the input. 
As the approach is deterministic, it simplifies the experimental preparation of topological states, preventing the need for fine-tuning the initial input.
While here, we exclusively considered the preparation of topologically protected modes, the concept of using a dissipation-free subspace to selectively filter states can be applied to any non-Hermitian system exhibiting a dark state. Generalizing this work to the quantum regime is a compelling next step, particularly by exploring the use of multi-photon light states across a wide range of established topological lattices that host protected zero modes. This passive and robust protocol therefore promises to advance integrated photonics by enabling the efficient and reliable on-demand generation of coherent and possibly topologically protected states. 

\section{Acknowledgments}
A.B.-R. acknowledges support by the National Science Foundation (NSF) award number 2328993.
A P.-L. is supported by MURI grant from Air Force Research Office (programmable systems with non-Hermitian quantum dynamics: FA9550-21-1-0202).
K.B. acknowledges funding by the German Research Foundation (DFG) in the framework of the Collaborative Research Center (CRC) 1375 (Project ID 398816777 -Project A06).
We thank Konrad Tschernig for his support in the early stages of this project.

\bibliographystyle{apsrev4-2}
\bibliography{bib_short.bib}
\end{document}